%% file: sample-acmsmall.tex
\documentclass[acmsmall,nonacm]{acmart}

\usepackage{xcolor}
\usepackage{pifont}
\newcommand{\cmark}{\ding{51}}%
\newcommand{\xmark}{\ding{55}}%

\AtBeginDocument{%
  \providecommand\BibTeX{{%
    \normalfont B\kern-0.5em{\scshape i\kern-0.25em b}\kern-0.8em\TeX}}}

\newcommand{\toolname}{\texttt{AccelMerger}} 

\definecolor{Crimsonglory}{rgb}{0.75, 0.0, 0.2}
\definecolor{applegreen}{rgb}{0.55, 0.71, 0.0}

\usepackage{fancyhdr}
\usepackage[normalem]{ulem}

\usepackage[normalem]{ulem}

\usepackage[T1]{fontenc}
\usepackage[utf8]{inputenc}
\usepackage[english]{babel}
\usepackage{todonotes}
\usepackage{textcomp}
\usepackage{xcolor}
\usepackage[detect-all]{siunitx}
\usepackage{comment}
\usepackage{algorithm}
\usepackage{algpseudocode}
\usepackage{balance}
\usepackage{mathtools}
\usepackage{float}
\usepackage{blindtext}
\usepackage{subfig}
\usepackage{pifont}
\usepackage{xcolor}
\usepackage{graphicx}
\usepackage{wrapfig}
\usepackage{mdframed}

\usepackage{listings}
\usepackage{etoolbox}

\definecolor{codegreen}{HTML}{54AA68}
\definecolor{codegray}{rgb}{0.5,0.5,0.5}
\definecolor{codepurple}{rgb}{0.58,0,0.82}
\definecolor{backcolour}{HTML}{E9EAF2}
\definecolor{keywordcolour}{HTML}{C54E50}

\definecolor{seabornblue}{HTML}{64B5CE}
\definecolor{seabornorange}{HTML}{DE8452}
\definecolor{seaborngreen}{HTML}{54AA68}
\definecolor{seabornred}{HTML}{c54e50}
\definecolor{seabornpurple}{HTML}{8274B4}
\definecolor{seabornbrown}{HTML}{947960}
\definecolor{seabornyellow}{HTML}{CDBA75}

\lstdefinestyle{mystyle}{
  backgroundcolor=\color{backcolour},   commentstyle=\color{codegreen},
  keywordstyle=\color{keywordcolour},
  numberstyle=\tiny\color{codegray},
  stringstyle=\color{codepurple},
  basicstyle=\ttfamily\footnotesize,
  breakatwhitespace=false,         
  breaklines=true,                 
  captionpos=b,                    
  keepspaces=true,                 
  numbers=left,                    
  numbersep=5pt,                  
  showspaces=false,                
  showstringspaces=false,
  showtabs=false,                  
  tabsize=2
}
\fancypagestyle{firstpage}{
  \fancyhf{}
  
  \fancyhead[C]{\vspace{15pt}\normalsize{MICRO 2020 Submission
      \textbf{\#\microsubmissionnumber} -- Confidential Draft -- Do NOT Distribute!!}}
  \fancyfoot[C]{\thepage}
}

\usepackage{calrsfs}

 \setcopyright{none} 
 \makeatletter
\let\@authorsaddresses\@empty
\let\@nonacm\@true
\makeatother

\makeatletter
\def\runningfoot{\def\@runningfoot{}}
\def\firstfoot{\def\@firstfoot{}}
\makeatother

\setcopyright{none}
\renewcommand\footnotetextcopyrightpermission[1]{}
\begin{document}

\title{Early DSE and Automatic Generation of Coarse Grained Merged Accelerators}


  

\author{Iulian Brumar}
\email{ibrumar@g.harvard.edu}

\affiliation{%
  \institution{Harvard University}
  \streetaddress{P.O. Box 1212}
  \city{Cambridge}
  \state{Massachusetts}
  \country{USA}
  \postcode{43017-6221}
}

\author{Georgios Zacharopoulos}
\email{georgios@seas.harvard.edu}
\affiliation{%
  \institution{Harvard University}
  \streetaddress{P.O. Box 1212}
  \city{Cambridge}
  \state{Massachusetts}
  \country{USA}
  \postcode{43017-6221}
}

\author{Yuan Yao}
\email{yaoyuannnn@gmail.com}
\affiliation{%
  \institution{Harvard University}
  \streetaddress{P.O. Box 1212}
  \city{Cambridge}
  \state{Massachusetts}
  \country{USA}
  \postcode{43017-6221}
}

\author{Saketh Rama}
\email{rvsaketh@gmail.com}
\affiliation{%
  \institution{Harvard University}
  \streetaddress{P.O. Box 1212}
  \city{Cambridge}
  \state{Massachusetts}
  \country{USA}
  \postcode{43017-6221}
}

\author{Gu-Yeon Wei}
\email{guyeon@g.harvard.edu}
\affiliation{%
  \institution{Harvard University}
  \streetaddress{P.O. Box 1212}
  \city{Cambridge}
  \state{Massachusetts}
  \country{USA}
  \postcode{43017-6221}
}

\author{David Brooks}
\email{dbrooks@g.harvard.edu}
\affiliation{%
  \institution{Harvard University}
  \streetaddress{P.O. Box 1212}
  \city{Cambridge}
  \state{Massachusetts}
  \country{USA}
  \postcode{43017-6221}
}


\begin{abstract}
\input{abstract}
\end{abstract}




\maketitle

\section{Introduction}

\input{intro}

\section{Related Work}
\label{sec:relWork}

\input{related}

\section{Problem Definition}
\label{sec:problem}
\input{problem}
\vspace{-0.2em}
\section{AccelMerger}
\vspace{-0.2em}

\input{methodology}

\subsection{Frontend}
\input{frontend}

\subsection{Function Merging}
\input{fmerging}

\subsection{HW/SW Partitioning}
\input{swhw}

\section{Experimental Results}
\input{experiments}

\section{Conclusions}
\input{conclusion}
\bibliographystyle{ACM-Reference-Format}
\bibliography{sample-acmsmall}










\end{document}

%% file: abstract.tex
Post-Moore's law area-constrained systems rely on accelerators to deliver performance enhancements.
Coarse grained accelerators can offer substantial domain acceleration, but manual, ad-hoc identification of code to accelerate is prohibitively expensive.
Because cycle-accurate simulators and high-level synthesis (HLS) flows are so time-consuming, manual creation of high-utilization accelerators that exploit control and data flow patterns at optimal granularities is rarely successful.
To address these challenges, we present AccelMerger, the first  automated methodology to create coarse grained, control- and data-flow-rich, merged accelerators.
AccelMerger uses sequence alignment matching to recognize similar function call-graphs and loops, and neural networks to quickly evaluate their post-HLS characteristics.
It accurately identifies which functions to accelerate, and it merges accelerators to respect an area budget and to accommodate system communication characteristics like latency and bandwidth.
Merging two accelerators can save as much as 99\% of the area of one.
The space saved is used by a globally optimal integer linear program to allocate more accelerators for increased performance.
We demonstate AccelMerger's effectiveness using HLS flows without any manual effort to fine-tune the resulting designs.
On FPGA-based systems, AccelMerger yields application performance improvements of up to 16.7$\times$ over software implementations, and 1.91$\times$ on average with respect to state-of-the-art early-stage design space exploration tools.

%% file: intro.tex




\label{sec:intro}

As CMOS scaling slows, and with it the increase in micro-circuitry per unit area, deciding which code regions to accelerate becomes harder. However, the demand for improved performance is exacerbated by novel domains such as AR/VR, Robotics and Machine Learning that require more agile and efficient design methodologies. The traditional approach in accelerating novel applications is to manually profile them and decide in an ad-hoc manner what code regions should be mapped onto hardware. These manual approaches are ad-hoc since they either directly map specific functions or loops onto hardware at a fixed granularity (e.g.~\cite{suleiman2019navion}) and/or informally identify a simple pattern occurring across different functions and create an accelerator for that (e.g.~\cite{chen2014diannao,pudiannao}). Code patterns can be as simple as repeated sequences of multiply-add operations to complex control and data flow graphs representing a function (CDFG) or a function call graph (nested-CDFG). In the case of manual accelerator design, patterns are only exploited at fine basic-block/DFG-level granularities, based on the architect's domain intuition.

To overcome some of these manual aspects of accelerator design, Early-Stage Design Space Exploration~\cite{zach2018regionseeker,zach2019compiler,kumar2016peruse} (Early-DSE) has emerged as a profiling methodology able to discover new SoC architectures adjusted to the application's characteristics without having to manually transform the most executed application functions into a synthesizable format, i.e. acceptable by High Level Synthesis Tools~\cite{legup,bambu,VivadoHLS,StratusHLSApr}, an extremely error-prone and time-consuming process. However Early-DSE comes unprepared to exploit variable CDFG granularities and much less to reason about architectures that reuse CDFG logic. 

Merging~\cite{venkatesh2011qscores,stitch,moreano2005efficient,cong2008pattern} small data-flow (DFG) circuits has been implemented with very-limited success using a graph-isomorphism approach, and at the RTL level, therefore missing an opportunity to think about the whole system in terms of reusing CDFG patterns. These approaches have not received widespread adoption in RTL synthesis tools~\cite{bambu}~\cite{legup}, most likely due to the fact that graph-isomorphism is NP-Hard and not very scalable, but also since at low granularities the benefits of calling an accelerator are greatly diminished by the cost of updating and reading from a register file (RF). In the HLS context, RF-less, coarse-grained accelerators have had the most success since rich CDFGs are mapped to hardware via datapaths handled via control-logic. There is no current technology able to merge these coarse grained accelerators.

\begin{figure}
  \centering
  \includegraphics[width=0.7\linewidth]{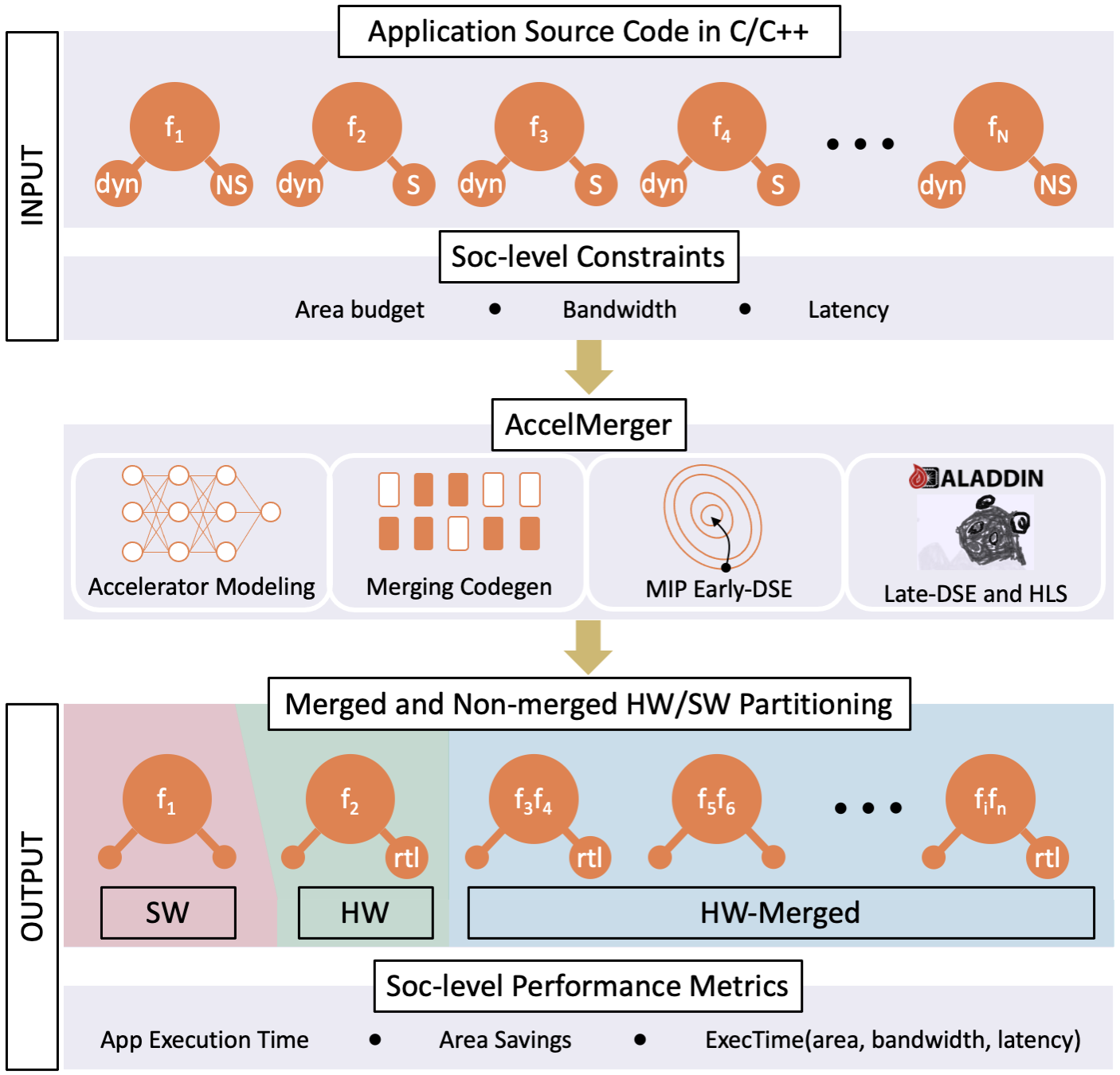}
  \caption{AccelMerger Overview.}
  \label{fig:accelMergerOverview}
  \vspace*{-2.0em}
\end{figure}%

In this paper, we present {\it AccelMerger}, an automated tool able to generate coarse-grained merged accelerators for the first time, to exploit CDFG patterns, able to generate new merged code regions with flexible granularities from loops and functions to entire call-graphs of merged and non-merged functions and to automatically select candidates for hardware acceleration. As illustrated in Figure~\ref{fig:accelMergerOverview}, AccelMerger starts from two pieces of information: the application source code containing potentially non-synthesizable functions as well as a configuration file specifying an area budget for a specific technology node and the parameters latency and bandwidth for the interconnect between the different accelerators and a general purpose CPU. If the area budget is not specified, or the interconnect latency/bandwidth, it will provide a scalability analysis for a wide range of relevant area budgets and bandwidths/latencies.

The four main steps performed in AccelMerger are illustrated in Figure~\ref{fig:accelMergerOverview}. \textbf{Accelerator Modeling}: We achieve  quick and automated accelerator modeling via Machine Learning Models, such as the Multilayer Perceptron (MLP). Our modelling relies both on the static information available in the application source code as well as on dynamic profiling information for each function indicated in Figure ~\ref{fig:accelMergerOverview} with the "dyn" label. This allows us to predict post-HLS accelerator resource consumption with less than 15\% error, which is very close to the error produced by late-DSE tools when applied to accelerator merging. \textbf{Merging Codegen}: Being able to quickly model accelerators goes hand-in-hand with code-generating merged functions and realizing them in hardware only if they present predicted opportunities for area savings, and by ensuring that the introduced multiplexing latency overheads do not cancel the benefit of allocating new merged and non-merged accelerators using the saved area resources.  For this we use a scalable sequence alignment approach to achieve merging at coarse granularities. \textbf{Late DSE and HLS}: We use cycle accurate simulation and HLS not only to create a dataset to predict accelerator area and latency but also to validate accelerators when the original functions are synthesizable.  \textbf{MIP Early-DSE}: When area wins are measured on the corresponding merged accelerators, we forward them to the selection stage via mixed integer linear programming (MIP) which determines the final list of merged and non-merged accelerators in the system. In particular for merged accelerators it is a requirement for the two input functions to be synthesizable to be able to run the final validation step via HLS and Cycle-Accurate Simulation. In Figure~\ref{fig:accelMergerOverview} we can see that functions such as $f_2$ are selected for acceleration and RTL can be generated for them as indicated by the "rtl" note attached to the function. The requirement for merged functions to be synthesizable however, is for both of the input functions to be synthesizable, which is the case for functions $f_3$ and $f_4$ but not for $f_i$ and $f_N$ since $f_N$ is non-synthesizable. AccelMerger does not require a function to be in a synthesizable form and might still suggest it for hardware acceleration since Neural Networks are used to perform accelerator modeling based on the instruction opcodes available in the high-level application. 

In experiments carried out on the SPEC CPU2006 benchmark suite \cite{henning2006spec} and the H.264 video decoder \cite{h264uiuc}, targeting Programmable Systems on Chip (PSoCs) Artix Z7007S and Artix Z-7012S \cite{xilinxPSOC}, we observe performance improvements of $1.91\times$ on average, with respect to state of the art Early-DSE, while remaining compatible
with well known, 
mature HLS toolflows.

This work makes the following contributions.\\

\begin{itemize}

    \item {\bf Merging of candidates for acceleration}. 
    To the best of our knowledge, we merge coarse-grained accelerators for the first time. We describe the machine learning models, the code generation techniques, the metrics, the dynamic profiling and the optimization techniques necessary to effectively identify and merge accelerators.
    \item {\bf Automated design space exploration for large-area designs}. We provide an automated toolflow from high-level C/C++ to RTL that generates merged accelerators when the application is synthesizable. When the application is only available in non-synthesizable format, AccelMerger is still able to provide insights about the code most amenable for acceleration and merging. By using Neural Networks, $4.5 \times 10^6$ merged accelerators can be analyzed in less than 10 seconds with AccelMerger, whereas pure Late-DSE techniques struggle to analyze eight hundred merged accelerators in less than 22 hours.
    
      \item {\bf Merged and non-merged accelerator selection}.
       AccelMerger is able to select the optimal mix of merged, non-merged, and software versions of the functions in the original program. We contribute a Mixed Integer Programming model that can operate on nested CDFGs while deciding what to merge based on an area budget.
\end{itemize}

To motivate our approach to DSE automation, we categorize in Section~\ref{sec:relWork} existing design tools and techniques for SoC accelerators. That leads to a statement (Section~\ref{sec:problem}) of the problem AccelMerger solves: efficient selection of application regions of varying granularity for implementation in hardware that is optimized for area constraints and chip characteristics. AccelMerger solves this problem (Section~\ref{sec:fmerging}) by searching for hot regions that match well enough to merge, saving area and allowing more accelerators to be allocated. Since performance improvements are derived from area savings and hardware realization of new accelerators, we describe in Section~\ref{sec:frontend} why Neural Networks provide increased accuracy as opposed to more explainable estimating models such as LASSO and Random Forests. In Section~\ref{sec:hw_sw_part} we describe the Linear Programming-based DSE that supports merging-aware accelerator selection and efficient scaling to the large numbers of candidates that merging introduces. Finally, Section~\ref{sec:setup} analyzes AccelMerger's design support for a variety of workloads, sweeping area budgets, granularities, interconnect latency and bandwidth.






%% file: related.tex
\renewcommand{\arraystretch}{1.3}
  \begin{table}[h]
  \Huge
  \centering
  \resizebox{0.7\linewidth}{!}{

    \begin{tabular}{|l|c|c|c|c|c|} 
\hline
  \textbf{Feature}    & \textbf{Fine grained}   &  \textbf{Manual} & \textbf{Early} & \textbf{Accel} \\
   & \textbf{Merging} & 
  \textbf{ DFG Merging } & \textbf{DSE} &  \textbf{Merger} \\
            \hline  \hline
  Application Time        &  \xmark    & \cmark  &  \cmark & \cmark \\ \hline
  Communication           &  \xmark & \cmark &\cmark  & \cmark    \\ \hline
  Fine Grained Merging                     &  \cmark  & \cmark & \xmark & \cmark  \\ \hline 
  \textbf{Coarse Grained Merging}                    &  \xmark   & \xmark & \xmark & \cmark  \\ \hline 
   \textbf{Early HW/SW Part.}               &  \xmark  & \xmark & \cmark   & \cmark \\ \hline
   \textbf{Flexible granularity}     &  \xmark  & \xmark &  \xmark   & \cmark \\ \hline
    \textbf{Automated}    &  \cmark  & \xmark &  \cmark   & \cmark \\ \hline

  \end{tabular}

  }  
  \vspace{0em}
  \caption{Taxonomy Table. We highlight the dimensions along which AccelMerger provides most contributions in \textbf{bold}.}
  \vspace{-0.4em}
  \label{tab:taxonomy}
  \end{table}

 Table~\ref{tab:taxonomy} presents a taxonomy of related work. The rows represent desirable features for accelerator merging/DSE tools and the columns represent bodies of related research. For each column, we look at the following features:
 
 \textbf{Application Time}. It indicates whether the set of techniques evaluate applications in an end-to-end manner instead of focusing on datasets of basic blocks as it occurs in the case of Fine grained DFG Merging.
 
 \textbf{Communication}. It indicates whether latency and/or bandwidth for the communication between the accelerators and the CPU is taken into account. The manual design of accelerators takes this into account using back-of-the-envelope calculations and eventually slow-cycle accurate Late-DSE~\cite{shao2014aladdin, shao2016co,salam}, that still requires manual application transformations, to validate a limited number of designs. Among prior Early-DSE techniques, only AccelSeeker~\cite{zach2019compiler} supports latency estimation. For synthesizable applications, in which the amount of data produced and consumed is easier to analyze statically, AccelMerger supports bandwidth estimation as well.
 
 \textbf{Fine grained merging}. In manual accelerator design, the architects might come up with Processing-Element-based accelerators which capture common DFGs relevant across an application. The fine grained DFG literature has automated this step on small basic-block datasets.
 
 \textbf{Early HW/SW Partitioning}. Other than the Early-DSE literature, manual accelerator design and fine-grained merging do not approach accelerator design in a unified and systematic manner tying together resource consumption with the performance maximization problem.
 
 \textbf{Coarse Grained Merging} and \textbf{Flexible Granularity}. To the best of our knowledge AccelMerger is the first to achieve these two goals.

 We next discuss the columns in Table~\ref{tab:taxonomy}:
 
\textbf{Early-DSE}: The only other Early-DSE related work to use Machine Learning as part of its pipeline is Peruse~\cite{kumar2016peruse}. Peruse however does not tie together resource area constraints with the system performance maximization as a single optimization problem. More recently, RIP~\cite{zuo2017accurate} and Accel/RegionSeeker~\cite{zach2019compiler,zach2018regionseeker} formulated the problem as a single optimization procedure using mixed integer programming and ad-hoc algorithms respectively. RIP is very limited by its applications, containing only two end-to-end applications and similarly AccelSeeker is only demonstrated on one application. Neither RIP nor AccelSeeker use machine learning to estimate key accelerator statistics. We observe that this is necessary for Early-DSE in general since the application CDFG goes through many complex transformations during HLS, and estimating accelerator statistics directly on the original application CDFG is challenging. An ML-approach is even more important for AccelMerger in order to effectively filter the huge number of merged accelerators.  

\textbf{Manual DFG Merging}: Noteworthy implemented accelerators that benefit from manually merging computational patterns include the PuDianNao and DianNao accelerators~\cite{chen2014diannao, pudiannao}, which are built specifically for machine learning workloads and reuse common computational patterns such as activation functions and typical linear algebra operators. AccelMerger is aimed at informing the architect about even more profitable accelerator merges at a coarse granularity. If the application code is available in synthesizable format, AccelMerger wiil also provide the RTL for the merged accelerators.

\textbf{Fine Grained Merging}: QsCores~\cite{venkatesh2011qscores}, is an infrastructure that generates accelerators tightly coupled to the CPU via the L1 cache and programmable through a specific interface that allows arbitrary control transitions, a model that is incompatible with many HLS and cycle accurate simulation flows. Similarly Stitch~\cite{stitch} is focused on generating such tightly coupled fine grained accelerators for wearable applications. While these two approaches together with older contributions focus on merging dataflow graphs available in basic blocks~\cite{brisk2004area,lam2009rapid, moreano2005efficient,venkatesh2011qscores,stitch}, AccelMerger is compatible with state-of-the-art HLS and cycle accurate simulators, since it supports the full complexity of a function, including control, data flow, and call graphs. \cite{brisk2004area,lam2009rapid, moreano2005efficient} represent a fine grained merging body of work focused on small basic block datasets. These approaches are designed for Late-DSE when the application fragments to accelerate have already been determined and transformed to make use of these fine grained accelerators. AccelMerger is completely designed for scalability and the early design stages. 

\textbf{Other} contributions relevant for AccelMerger: For Late-DSE techniques it is also worth mentioning Hetero-CL~\cite{heterocl}, Spatial~\cite{koeplinger2018spatial} and Aetherling~\cite{aetherling}. Even though they present significant performance, accuracy vs performance tradeoffs and programmability advantages when compared to writing Verilog or VHDL, the architect would still have to port an application to a specific DSL (domain specific language) to benefit from the DSE available in these tools. Moreover, these tools do not exploit accelerator merging.

FMSA and SalSSa~\cite{rocha2019function,rocha2020effective} were recently introduced as LLVM~\cite{lattner2004llvm} compilation transformations that target code reduction for embedded devices. Work prior to FMSA is only able to merge equal functions whereas FMSA is able to merge functions with different argument lists, returned values and references as well as differing control flow. FMSA does not take into account the dynamic behavior of the application nor does it try not to hurt the application performance by introducing less multiplexing in the application hot spots. Finally, neither FMSA nor other function merging approaches have been used for accelerator merging. Both the function merging and the Early-DSE related-work~\cite{rocha2019function,rocha2020effective,kumar2016peruse} demonstrate their results on the SPEC CPU2006 suite, and we include these applications in this paper for comparison.

    


%% file: problem.tex
\begin{figure}

  \centering
  \includegraphics[width=0.7\linewidth]{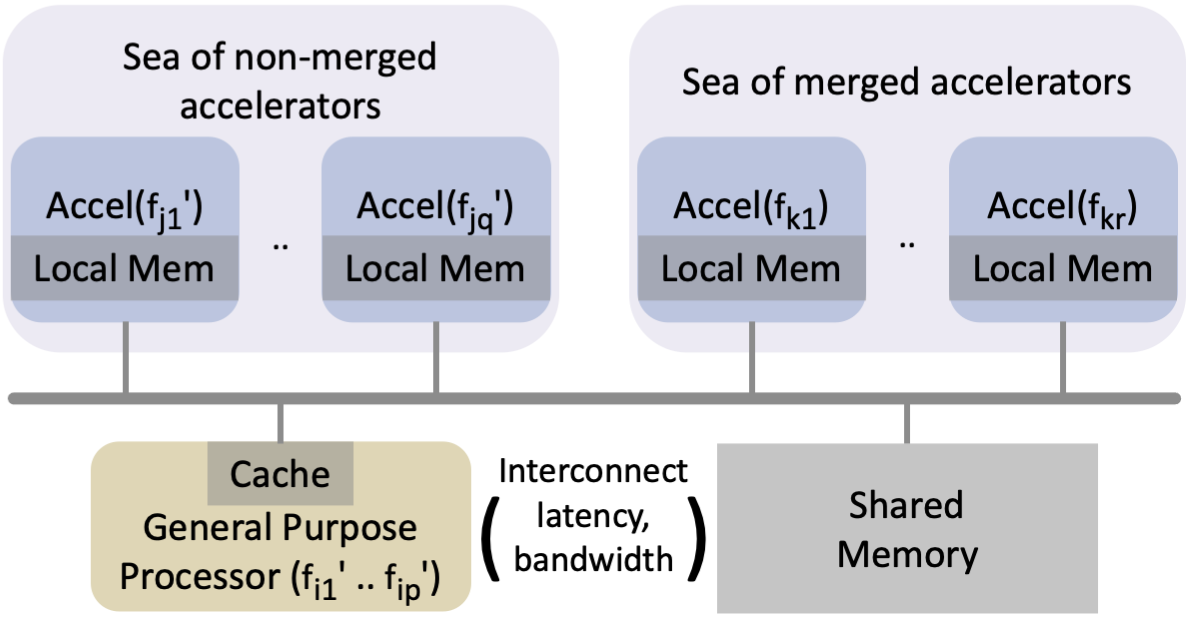}
  \caption{AccelMerger's SoC model and feedback template for the system architect. This is a lower level view of the Merged and Non-merged accelerator selection output in Figure~\ref{fig:accelMergerOverview}.}
  \label{fig:systemModel}
  \vspace*{-0.5em}
\end{figure}%

\textbf{Vanilla HW/SW Partitioning}. We first describe the vanilla HW/SW partitioning problem as addressed in the related work and then proceed to the larger problem present in systems with limited area resources. Given a set of functions $P=\{f_1, .. , f_n\}$, each best represented by a CDFG, a list of arguments and a returned type, the vanilla HW/SW partitioning generates two sets of functions $A = \{f_{i_1}, .., f_{i_p}\}$ and $B = \{f_{j_1}, .., f_{j_q}\}$ to execute on a general purpose core and on specialized hardware, respectively. The two sets represent a partitioning of $P$ since $A \cap B = \emptyset$ and  $A \cup B = P$. This partitioning is done by maximizing program performance subject to the total resource requirements of the accelerators and the general purpose CPU, along with a description of how these components communicate. The related work in early-stage DSE~\cite{zach2018regionseeker,zach2019compiler} operates on parameters $sw_i$, $hw_i$ and $area_i$, which are automatically determined for each function. $sw_i$ and $hw_i$ represent the overall time spent in seconds executing a function in software, using a general-purpose core and in hardware, using coarse-grained accelerators, respectively. $area_i$ is the area requirement per component measured in Lookup Tables (LUTs). Some uncommon features that are usually only modeled in late-stage DSE tools, such as gem5-aladdin~\cite{gem5aladdin} and gem5-salam~\cite{gem5salam}, are the interconnect $latency$ and $bandwidth$ that provide insights about the system-level, and we take these into account when determining the optimal partitioning. In contrast with late-stage DSE, which is focused on evaluating point-solution SoC designs with the predefined set of functions to accelerate and run in software, we need an agile way of determining which loops and functions to accelerate and merge. 



Moreover, for Post-Moore's law accelerators with limited compute fabric, we need to build an extended set of functions that result from merging the most similar functions in $P$. These would enable hardware realization with lower area cost. The creation of merged accelerators needs to overcome three technical issues, described in the following paragraphs.

\begin{figure}[t]
\centering
\includegraphics[width=0.7\linewidth]{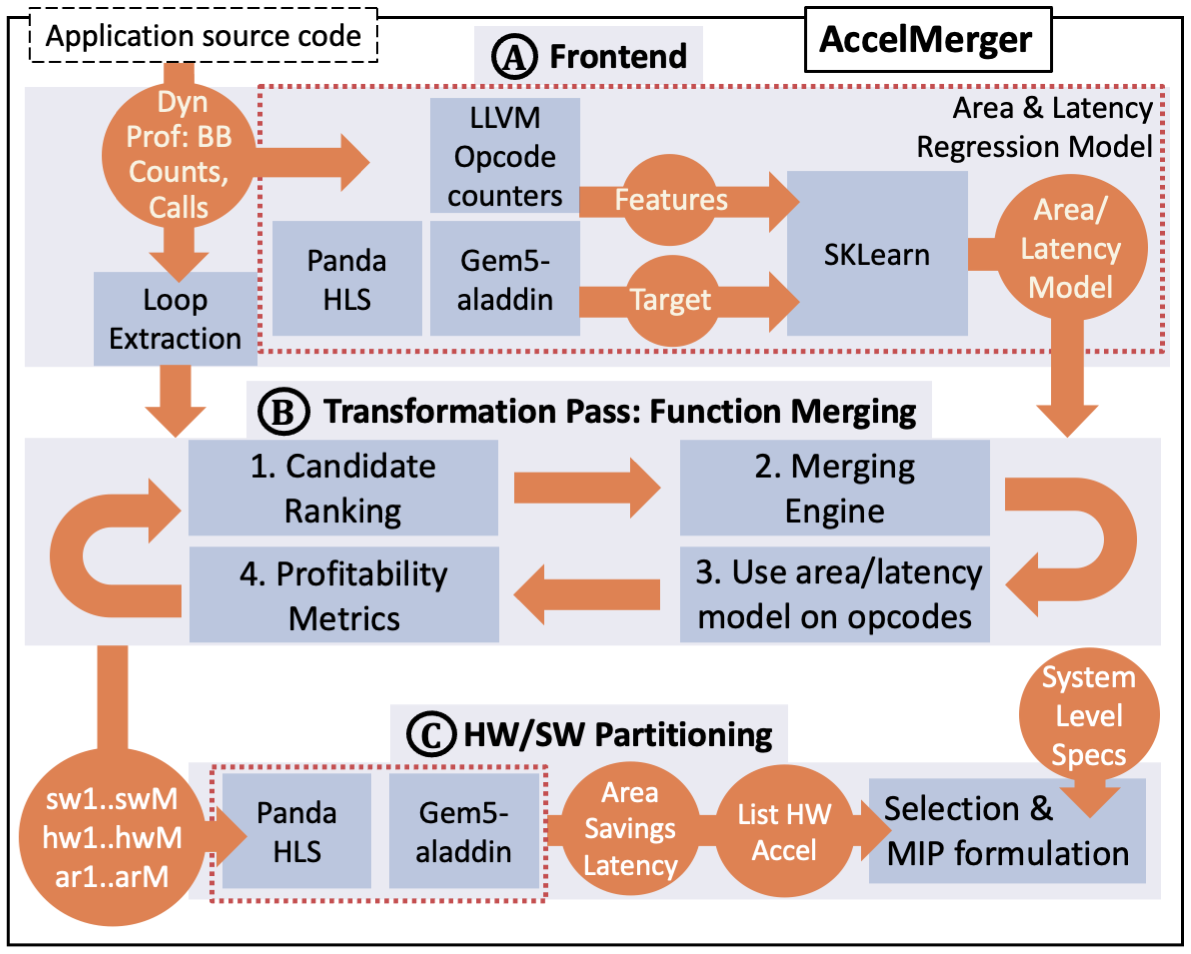}
\caption{\toolname's workflow. A) Accelerator Modeling and Loop Extraction. B) Function Merging Pass. C) HW/SW Partitioning. }
\vspace*{-0.5em}
\label{fig:workflow}
\end{figure}

\textbf{Function Merging at arbitrary granularities}. We need to expand the set $P$ with merged functions $M=\{f_{k_1}, .. , f_{k_l}\}$, that exploit flexible computational granularities and CDFG patterns to produce a new set $P'=P \cup M$, containing new merged functions and functions from the original set $P$. When merging two "parent" functions, $f_{i_1} \in P$ and $f_{i_2} \in P$, the desired effect is that the result, say $f_{k_1}$, satisfies $area_{f_{k_1}} < area_{f_{i_1}} + area_{f_{i_2}}$. The exact code transformation to create these new candidates, namely \textit{Function Merging}, should do more than just concatenate the CDFGs of the parent functions. Instead, it should reuse nodes corresponding to instructions in the high-level functions, and to RTL functional units at the lower level if these functions are synthesizable. We can express the Function Merging transformation more concisely as
\vspace{-5pt}
\begin{align*}
\textrm{fm}:P \times P &\rightarrow M \\
(f_{i} ,\; f_{j}) &\mapsto f_k
\end{align*}
 The size of $M$, denoted $|M|$, is less than or equal to $n^2$ depending on how many similar functions there are in $P$, therefore $n + 1 \leq k \leq n + 1 + n^2$. We use subindexes starting from $n+1$ for $k$ to avoid confusing the $f_k$ functions in $M$ with functions in $P$ with subindexes $1..n$. We address the CDFG reuse problem in Section~\ref{sec:fmerging}. 

\textbf{Accelerator Modeling}. Since the set of possible merged functions $M$ grows quadratically with the number of functions and loops in the original program $P$, Early-DSE needs to be able to model $area_i$ quantities with high accuracy since area reduction is the main criterion for choosing to merge two functions. For hardware and software latency, the related work has indicated that simple linear models can provide estimates for $sw_i$ and $hw_i$ that result in overall high quality feedback about the most desirable functions to accelerate~\cite{zach2019compiler}. In Section~\ref{sec:frontend}, we describe how AccelMerger is able to model area reduction for merged accelerators with high accuracy.

\textbf{HW/SW Partitioning}. In the context of the new program $P'$, HW/SW partitioning is significantly complicated because $P'$ contains functions that have overlapping functionalities, and because RTL flows create hardware for a function $f$ in a hierarchical manner, including the CDFGs of $f$'s callees. These interactions include the fact that merged functions still need to be considered in the context for $P$'s call graph. For example, if a merged function is realized in hardware, its callees need to be realized in hardware as well. Solving the HW/SW Partitioning problem with merged accelerators will result in an output as illustrated in Figure~\ref{fig:systemModel}. The HW/SW partitioning step needs to produce three sets $A'= \{f_{i_1}', .., f_{i_p}'\}$, $B'=\{f_{j_1}', .., f_{j_q}'\}$ and $C = \{f_{k_1}, .., f_{k_r}\}$ such that $A' \subseteq P$ and $B' \subseteq P$ and $C \subseteq M$. $A$, $B$ and $C$ cannot be considered a partition in the mathematical sense since there can be more functions in $P'$ than in the union $A \cup B \cup C$, meaning that before the HW/SW partitioning step the function merging step might generate merged functions $f_k$ that are not selected in the final SoC when the parent functions $\textrm{fm}^{-1}(f_k) = (f_i, f_j)$ are more profitable for a specific area budget and SoC communication characteristics.

We use $f_{i_1}'$ in contrast to $f_{i_1}$ to denote that for a given area budget, the functions we will decide to have in software and hardware will differ from these selected and generated by vanilla Early-DSE. Subset $C$ represents merged functions that are mapped onto hardware but never on to the general purpose software processor, since merged functions present benefits for accelerator reduced area consumption which can be leveraged by the accelerators only.   In Figure~\ref{fig:systemModel}, we observe that the interconnect we model between the CPU and the accelerators has latency and bandwidth properties which is information the architect can provide in a configuration file as indicated in Figure~\ref{fig:accelMergerOverview}. For each call to a function that is mapped onto hardware, it takes "latency" cycles to initiate the computation plus the time required to move data from the CPU or another accelerator's local memory according to the available bandwidth. We indicate how AccelMerger handles this larger HW/SW partitioning problem in Section~\ref{sec:hw_sw_part}.





%% file: methodology.tex
\textbf{\textcircled{A} Frontend:} Accelerator Modelling and Loop Extraction are described in Section~\ref{sec:frontend}. The input to this stage is the application source code and the main result is a model for coarse grained accelerator area prediction. Also the application is transformed to also enable loop-level merging. \textbf{\textcircled{B} Function Merging} in Section~\ref{sec:fmerging} is a transformation pass used to merge functions taking into account their characteristics when mapping them onto hardware using the model from step \textcircled{A} and latency prediction models from the related work. In order to determine whether the merged functions make sense in the context of the system specification, both the new merged functions and the original functions are forwarded to the final step. \textbf{\textcircled{C} HW/SW partitioning} is the step where we determine the final layout of the application in terms of functions to be executed in software, in merged hardware and non merged hardware depending on the available area budget, and interconnect characteristics. This step is described in Section~\ref{sec:hw_sw_part}. Finally, Section~\ref{sec:setup} analyzes AccelMerger's design support for a variety of workloads, sweeping area budgets, and interconnect latency and bandwidth.




%% file: frontend.tex

\label{sec:frontend}
\label{sec:modelling}

As detailed in Section \ref{sec:fmerging}, merging needs to quickly evaluate the feasibility of a merge operation as the number of possible merges increases quadratically.\footnote{The complexity can be larger since merged functions can be merged again in high-similarity programs.} 

\textbf{Area and Latency Models}: For AccelMerger the most critical components that need an accurate evaluation are the coarse grained accelerators that we will synthesize using Bambu HLS and simulate with the cycle-accurate trace-based simulator Aladdin. Recently, research has focused on the mapping classification problem, studied in the context of CPUs and GPUs. This code mapping problem is about determining which platform should execute a piece of code. \cite{optComp} and \cite{poem} have used Recurrent Neural Networks and Graph Neural Networks for this classification problem for GPUs and CPUs. However, there has been limited work on accurately predicting FPGA and ASIC accelerator statistics without integrating these predictions in an Early DSE infrastructure~\cite{kumar2016peruse}. Tools such as gem5-aladdin or gem5-salam do not emphasize how latency models have been constructed as long as they achieve high accuracy, and most likely these models rely on architect's intuition about how fast an instruction typically runs. AccelMerger takes a mixed approach by modeling area in great detail, since we need to quickly determine whether there are area wins for many function pairs and discard less profitable candidates, and we use Aladdin's estimations for instruction latency, which have been shown to have an error as low as 0.9\%~\cite{shao2014aladdin}. Modeling the area consumption with our own Machine Learning models is further motivated by the fact that Aladdin models area consumption using a linear model directly attributing a latency estimate to each LLVM instruction, resulting in a $7.22\times$ higher error when measuring area than for latency.

\renewcommand{\arraystretch}{1.2}
  \begin{table}[h]
  \centering
  \resizebox{0.6\linewidth}{!}{
    \begin{tabular}{|l|c|c|c|c|c|} 
\hline
  \textbf{Accuracy vs}    & $r^2$ train  &  $r^2$ test & MRE Train & MRE Test \\
   Model &  & &  &  \\
            \hline  \hline
  LASSO-200        &  0.82    & 0.67  &  0.44 & 0.49 \\ \hline
  Random       &  0.96 & 0.86 & 0.12  & 0.37  \\
  Forest-200      &   &  &   &   \\ \hline
  MLP-200           &  0.99 & 0.9 & 0.1  & 0.29    \\ \hline \hline
    LASSO-600        &  0.96    & 0.95  &  0.24 & 0.35 \\ \hline 
  Random       &  0.98 & 0.89 & 0.12  & 0.27  \\
  Forest-600      &   &  &   &   \\ \hline
  MLP-600           &  0.99 & 0.97 & 0.15  & 0.21    \\ \hline
  \end{tabular}

  }  
  \vspace{0em}
  \caption{Model Selection and Accuracy table for area wins prediction. LASSO stands for the "Least Absolute Shrinkage and Selection Operator" linear prediction model and MLP stands for the Multilayer Perceptron, a neural network model.}
  \label{tab:modelAccuracy}
  \end{table}

\begin{figure}
  \centering
  \includegraphics[width=0.7\linewidth]{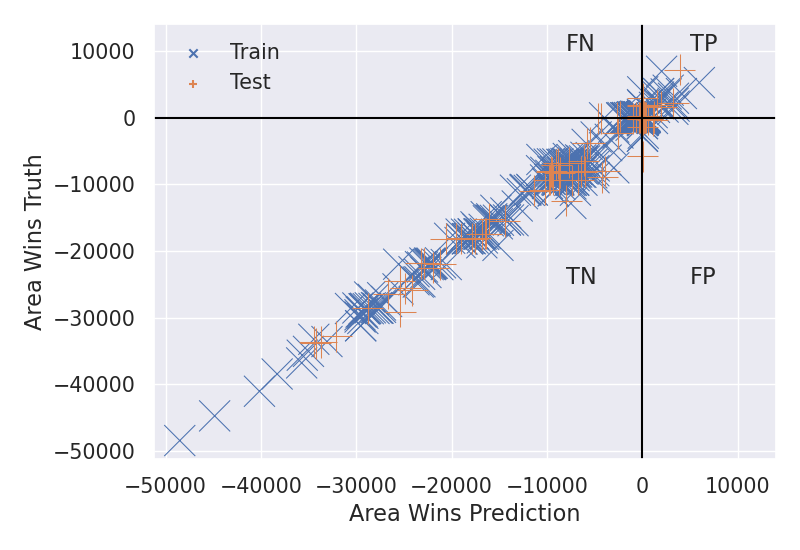}
  \caption{Multilayer Perceptron (MLP-600) area model for LUT wins. Synthesized/Real area vs. Predicted Area. TP = True Positive, FP = False Positive, TN = True Negative, FN = worse case scenario, accelerators not produced which would have brought area benefits}
  \label{fig:areaModelValid}
\vspace*{-0.5em}
\end{figure}%

\textbf{Features used to predict Area and Latency}: We model the area required for each static instruction using multiple techniques including Multilayer Perceptron, Random Forests and the LASSO linear model. The features we use are the number of LLVM operations of each type and the base truth HW resources number, obtained with Bambu HLS for area. When modeling the area for a function $f$, we count the LLVM instruction features hierarchically, taking into account the static number of instructions for $f$'s callees. We obtain the models by training on both merged and non-merged synthesizable functions in H.264 and MachSuite~\cite{reagen2014machsuite}. We configure Bambu HLS to allocate logic by using Lookup Tables (LUTs) exclusively to simplify area consumption and speedup comparisons across different approaches. In order to estimate the accelerator latency we use Aladdin's latency model on the hierarchical counts of dynamic instructions for each instruction type.

\textbf{Model Hyperparameters}: Table~\ref{tab:modelAccuracy} shows the result of predicting the area wins for merged accelerators using different machine learning models. We will explain in Section~\ref{sec:fmerging} the exact mechanism that allows us to produce merged accelerators and in this section we focus exclusively on how we measure their area wins. The accuracy numbers shown in the table are generated for the best model across an exhaustive (grid) search of model hyper-parameters, using 3-fold cross-validation. LASSO is the simplest model that we experiment with here and we tune 2 hyper-parameters, including the model complexity penalization alpha parameter, with a total of 6 combinations. For Random Forests we perform a grid search for trees with a specific maximum depth, a maximum number of estimators and a maximum number of features. The total number of hyper parameter configurations we explore for Random Forests is 360. For the Multilayer Perceptron (MLP) neural network, we tune the number of hidden layers, the number of neurons per layer, the activation functions, the complexity penalization parameter alpha, and the random seed. The total number of combinations explored is 360 and the best selected model had 6 hidden layers with 40 neurons each, ReLU activation function and $alpha=0$. This architecture was better than larger ones which were over-fitting the data and resulting in lower accuracies. 

\textbf{Accuracy Metrics}: Table~\ref{tab:modelAccuracy} shows the model accuracy both for training each model on fewer data points (functions), 200 specifically in the case of the <model>--200 rows, as well as on more functions, 600 specifically in the case of the <model>--600 rows.  In the columns we display two common regression accuracy metrics. First, $r^2 = 1 - \frac{SS_{res}}{SS_{tot}}$ which is interpreted as the percentage of variance in the true target variable $y$ explained by the model $f$, with $SS_{res} = \sum_{i=0}^n (y_i - f_i)^2$ and $SS_{tot} = \sum_{i=0}^n (y_i - average(y))^2$.  We see that the LASSO linear model can generalize very well, provided that enough data is available. With only 200 data points, $r^2$ is of only 67\% though whereas the non-linear models can quickly model the patterns resulting from the multiple transformations HLS tools perform to transform high level code to RTL, even with less data points. $r^2$ is a loose accuracy metric even though popular in the related work. 

Secondly, we also measure the Mean Relative Error $MRE = \sum_{i=0}^n \frac{abs(y_i - f_i)}{abs(y_i)}$. This is the metric related work in Late-Stage DSE such as Aladdin~\cite{shao2014aladdin} uses to measure accuracy for area and latency. This is a much stronger metric and late-stage DSE is able to minimize this error by iterating over representative applications with dozens of optimizations over the dynamic data-dependence graph (DDDG) representative of HLS flows that generate hardware. In a sense, the related-work for late-DSE is reporting the equivalent of train-time MRE error, since models are created on the same data that is later reported as indicative of the in-production behavior. It is very remarkable that with the model MLP-600 we are able to achieve roughly the same MRE train error as Aladdin, but we also provide the MRE test error which is of 21\%. This is very remarkable since MLP is able to produce this accuracy just starting from information available in the high-level LLVM-code (the instruction opcodes), and thus, we're able to use this model on both synthesizable as well as non-synthesizable applications.


\textbf{Choosing the MLP model}: Figure~\ref{fig:areaModelValid} shows these results via the relationship between the predicted area savings, using the MLP-600 model, measured in LUTs and the real area savings that can be observed by synthesizing the merged accelerators. The figure demonstrates that we are effectively able to filter merges that are not profitable and more importantly it does not omit profitable merges. The figure also conveys the fact that that in the absence of an accurate area predicting model like MLP-600, it would be very hard to filter merged accelerators with area losses.

\textbf{Transforming Loops into Functions} One of the transformations in the LLVM infrastructure is described next. It enables merging at different granularities than function level only.



Merging accelerators exclusively at function level yields different degrees of performance depending on how well the code has been modularized. To overcome this issue, AccelMerger's frontend creates new functions that correspond to loops in the original code. Our infrastructure can be run in two modes: a) operating on the functions present originally in the application, or b) converting outer loops into functions and replacing them with function calls. The former has the disadvantage of fewer merging opportunities, but it avoids the extra calling and parameter passing overhead of working with loop extraction. To achieve the latter, we use the \emph{LoopExtraction pass} available in LLVM. The pros and cons of loop extraction are discussed extensively in Section~\ref{sec:setup} and in \cite{kumar2016peruse}.




%% file: fmerging.tex
\label{sec:fmerging}



\begin{figure}
  \centering
  \includegraphics[width=0.7\linewidth]{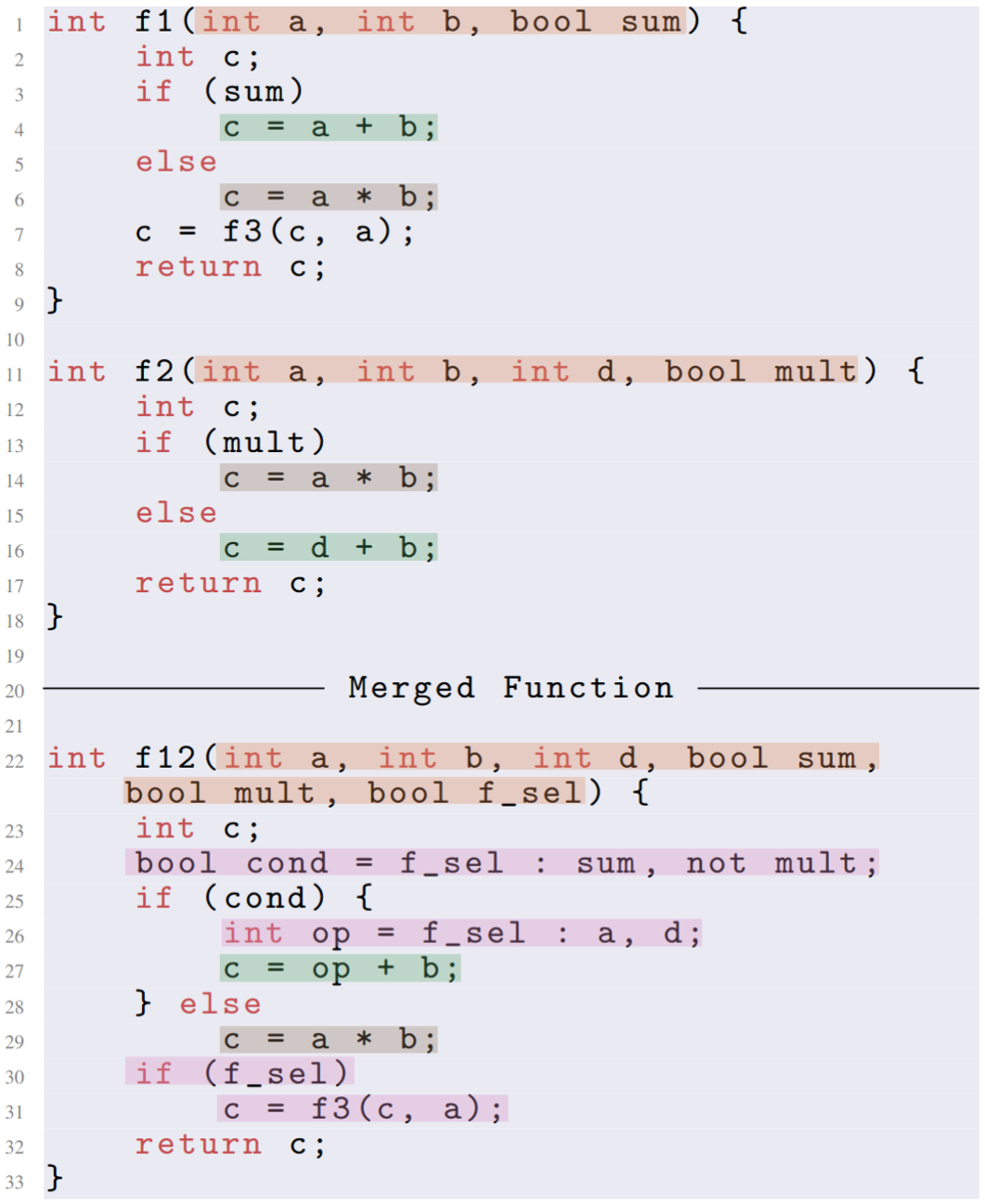}
  \caption{Simple function merging example expressed in high level C-like pseudo-code.}
  \label{fig:funcMergingExample}
  \vspace*{-0.5em}
\end{figure}%

AccelMerger starts from a high-level application and places accelerator latency and area benefits in the context of the full application~\cite{shao2016co}. It takes advantage of the flexible LLVM representation to produce for the first time, to the best of our knowledge, merged accelerators of arbitrarily large sizes, constrained of course by the amount of hardware area available. Step \textcircled{B} Transformation Pass: Function Merging,  in Figure~\ref{fig:workflow} uses the area prediction model generated in the Step \textcircled{A} to generate merged functions that when mapped onto hardware with a smaller resource consumption than the input functions. All the call sites for the input functions are not yet replaced with calls to the merged functions and this is deferred to step \textcircled{C} HW/SW Partitioning which will determine the final set of merged and non merged functions to accelerate and only calls to the selected merged functions will be replaced with merged accelerator invocations. 



\subsubsection{Function Merging Example}
Figure~\ref{fig:funcMergingExample} illustrates two simple similar functions that we can transform into a new function that is semantically equivalent to both input functions by using a flag $f\_sel$ that enables multiplexing non-aligned instructions and operands. In this section we only discuss how a function merging is realized provided that two functions look similar enough according to a simple heuristic we will later discuss. Once the merge is performed, we can apply the models described in Section~\ref{sec:frontend} on the input functions and the merged function to determine with high precision whether the merge is profitable. In this example we discuss how merging is being performed by following the methodology in \cite{rocha2019function} with some differences that make Function Merging amenable for accelerator merging and system design instead of code compression, which is the focus of the Function Merging related work~\cite{rocha2019function}~\cite{rocha2020effective}.

\textbf{Parameter merging} Figure~\ref{fig:funcMergingExample} shows two input functions that have some similarity regarding the input parameters. Function Merging starts by merging parameters depending on their types, not depending on their names. A direct approach is being taken in~\cite{rocha2019function} where for each parameter, in declaration order, in the first function, a parameter of the same type is being searched, in declaration order, in the second function. If a match is found, we proceed with trying to match new parameters and if not, the currently analyzed parameter in the first function will not be merged. Not merging a parameter means that whenever it is encountered as an operand to a merged instruction, it will be multiplexed using a $select$ instruction and the $f\_sel$ flag. For the second parameter in f1, we proceed with the remaining parameters in f2. ~\cite{rocha2019function} has shown this parameter merging approach to be very effective by intentionally modifying parameter merging taking into account the function body, and we observe similar area benefits in exploiting parameter merging of about $7\%$. 

\textbf{Merging function bodies}. We see in Figure~\ref{fig:funcMergingExample} that f1 and f2 have important similarities. The aligned instructions are determined using the Needleman-Wunsh (NW) sequence alignment algorithm~\cite{needleman} on the opcodes of the instructions. Since the algorithm NW operates on two strings and aligns their characters by finding the longest common subsequence (LCS) \footnote{NW is actually more complex than LCS since it allows to attach weights to the most important instruction types, a very important feature for accelerator driven function merging.}, a linearization preprocessing of both f1 and f2 is necessary since at the IR level, code is represented as a CDFG. NW indicates which instructions can be reused and which ones cannot.  In this simple example only slight discrepancies must be settled such as the differing condition for the $if$ statement at line 24, the discrepancy for the first operand of the addition operation at line 26 and the fact that f1 has a call not present in f2 at lines 30 and 31. These discrepancies are resolved via $select$ instructions to select the right operands and via $if$ statements for discrepancies not related to operand selection but whole instructions or groups of instructions that were not aligned across the two functions. For example the $call$ of f3 at the end of f1 is not present in f2 and therefore is only executed when f12 is called from f1's call sites.

\subsubsection{Accelerator Driven Function Merging}

We now describe how we systematically approach the merging problem in step \textcircled{B} box 1 of Figure~\ref{fig:workflow}, "Transformation Pass: Function Merging".

\textbf{1. Candidate Ranking}. We start the function merging process by ranking function pairs according to simple fingerprints that indicate how many shared instructions there are between two functions. This feature enables merging to scale to large code bases. Then the most similar candidates are linearized, meaning that we convert the graph structure into a sequential string of instructions. In AccelMerger we sample different post order (each basic block is visited after all of its descendants) linearizations in search for a high similarity match. On each of the alignments we apply the NW algorithm that produces the list of aligned instructions used in the Merging Engine step. 





\textbf{2. Merging Engine}. For step \textcircled{B} box 2 in AccelMerger's workflow in Figure~\ref{fig:workflow}, we generate code for the merged function for almost all pairs of functions. We filter some function pairs at this step that align a statistically insignificant number of instructions (i.e. less than $5\%$). The merging engine handles mismatching parameters and unaligned IR instructions by introducing overhead branch instructions and therefore creating new basic blocks for large chunks of unaligned code and for individual unaligned instructions, select instructions are used as multiplexors for the results of unaligned instructions with $f\_sel$ as a control flag. Parameters are handled as described earlier in this section. The most important piece of information used in this step is the alignment computed in the previous Candidate Ranking.  

\textbf{3. Using area/latency model on opcodes} Next, in step \textcircled{B} box 3 we use the area-predicting MLP-600 model presented in Section~\ref{sec:modelling} to predict with high accuracy the LUT consumption and the Aladdin per-instruction model.  If MLP-600 predicts the resource consumption using the opcode counts for the merging input functions and the opcode counts of the merging result.

\textbf{4. Profitability Metrics}. In Step \textcircled{B} box 4 we filter the merged functions that do not have area wins and the functions with unacceptable latency overhead. In this step we determine if the area of the merged function $f_{k_1}$ is smaller than the area of the input functions $f_{i_1}$ and $f_{i_2}$ altogether (i.e. $area_{f_{k_1}} < area_{f_{i_1}} + area_{f_{i_2}}$). If a merge passes this test, we check if the resulting accelerator is acceptable latency-wise.

\label{sec:profMetric}

For each merging input functions we denote their corresponding hardware latencies $hw_1$ and $hw_2$, their software latencies $sw_1$ and $sw_2$, and their area consumption $area_1$ and $area_2$. The resulting merged function is denoted with specifications $sw_{12}$, $hw_{12}$, and $area_{12}$. In order to establish large benefits for the merged accelerator, it is necessary that, when there is enough area for the merged accelerator only, but not for the two input accelerators, the merged accelerator can bring larger improvements than hardware realization of either of the input functions. Equation~\ref{eq:maxFuncProf} describes the maximum Estimated Profitability (EP).
\begin{equation}
    \label{eq:maxFuncProf}
    EP = \frac{sw_1 + sw_2 - hw_{12} - max(sw_1 - hw_1, sw_2 + hw_2)}{Total\_Application\_Exec\_Time}
\end{equation}

A profitable EP score is positive and the key insight that makes this equation area agnostic and useful is that, provided enough area to realize a merged accelerator in hardware, there also must be enough area available to realize the most profitable of the two parent accelerators. The maximum time savings benefit corresponding to the input functions is $max(sw_1 - hw_1, sw_2 - hw_2)$ and the benefit corresponding to the time saved by the merged function is $sw_1 + sw_2 - hw_{12}$. We only consider functions with positive EP as second filter after the area reduction check, to further narrow down the candidates for acceleration.

In the next section we describe the final selection of merged, non-merged and software-executed functions.

%% file: swhw.tex
\label{sec:hw_sw_part}



For the hardware/software partitioning formulation we use Mixed Integer Linear Programming (MIP). We stay consistent with our previous notation $area_i$, $hw_i$, $sw_i$. These numerical constants are the result of the ML-based modeling phase and indicate what would be the area and hardware execution latency of a function/loop if they were realized in hardware and $sw_i$ being the latency of executing each function/loop on the CPU.

\textbf{Software and hardware selection}. The MIP solver determines whether to realize a function in software or hardware by using the binary variables $hwv_i$ and $swv_i$. These variables are mutually exclusive. The objective function in Equation~\ref{eqn:obj} includes these two variables and their associated constant costs in latency, as well as a communication minimization term. It accounts for the latency and data transfer cost for transitioning from a software execution to a hardware one. The $\mathit{frontier}_{ij}$ variable will be 1 only when, $f_i$ is realized in software and $f_j$, a callee of $f_i$, is realized in hardware according to Equation~\ref{eqn:comm}. The amount of area for hardware acceleration is capped by Equation~\ref{eqn:max_area}.


\textbf{Handling call graph}. In the selection problem we operate both on the functions in the original program and on the tree of merged functions. To model the hierarchical aspect of coarse-grained accelerator generation in HLS tools, we take into account the program call graph to include all the functions called directly and indirectly by each function. 

\begin{figure}
  \centering
  \includegraphics[width=0.7\linewidth]{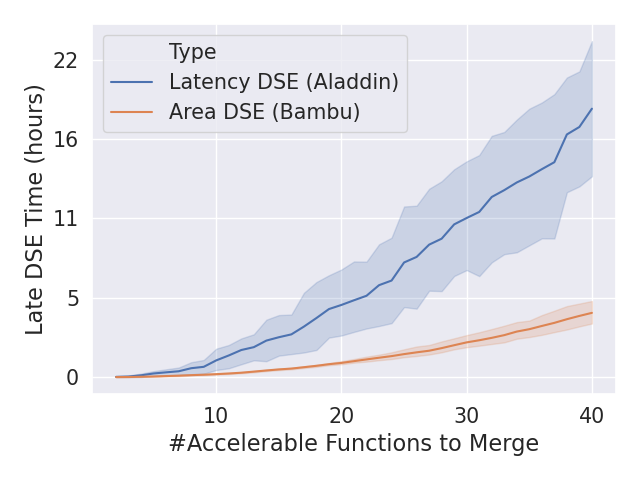}
  \caption{Hours spent in Aladdin to evaluate all possible merging candidates to determine their latency and in Bambu HLS to evaluate their area. The number of input functions for merging is represented in the x-axis of the figure. For 40 Functions to Merge, there are approximately $40^2/2 = 800$ merged accelerators to evaluate for all the possible pairwise merges. Late-DSE techniques struggle to analyze the possible merges of 40 accelerators in less than 22 hours. }
  \label{fig:aladdinDSE}
  \setlength{\belowcaptionskip}{-5pt}
  \vspace{-0.5em}
\end{figure}%

We consider a set $C_i$ of direct and indirect callees for each function. For each caller-callee pair of functions $f_i$ and  $f_j$ there is an associated constant number of dynamic calls from $f_i$ to $f_j$ "$calls_{ij}$" determined with dynamic instrumentation. 

\textbf{Handling merge graph}. When two high similarity functions $f_i$ and $f_j$ are merged, a new function $f_k$ is produced denoted the $child$ of the $parents$ $f_i$ and $f_j$. Since children themselves can be merged with other functions, a \textit{descendant} is a function obtained by repeatedly proceeding from parent to child.  For each function of an application we define a set $Descend_i$ that contains the children and other recursive descendants of function $f_i$. For a given $f_i$, $swv_i$ and $hwv_i$ can both be 0, meaning that $f_i$ is not implemented in software or in hardware, since one of its descendants $f_j$ might have $hwv_i = 1$. Similarly a function $f_k$ might not be realized either in software or in hardware since its ancestors are selected for the final SoC either in software or in hardware. We will call functions without parents \textit{Root} nodes.


Equation~\ref{eqn:appRealization} requires each Root to be realized either in hardware or software. Otherwise its functionality is covered by one of its descendants. Moreover if a function is realized in hardware, all its callees need to be realized in hardware as well but allowing the callee's functionality to be covered in hardware by its descendants, as shown in Equation~\ref{eqn:callAndMerge}. 

\begin{figure}
  \centering
  \includegraphics[width=0.7\linewidth]{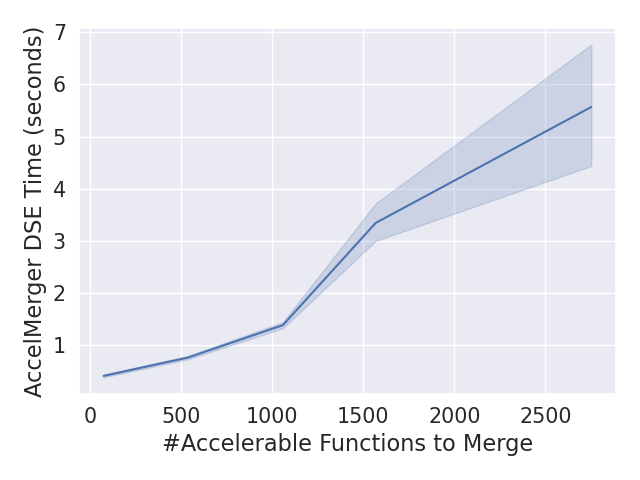}
  \caption{Seconds spent in AccelMerger performing DSE for a large range of functions. For 3000 functions  to  merge,  there are approximately $4.5*10^6$ possible  merged accelerators to evaluate for all the possible pairwise merges. Using accurate accelerator modelling with MIP-based DSE dramatically reduces analysis time. }
  \label{fig:accelMergerDSE}
  \setlength{\belowcaptionskip}{-5pt}
  \vspace{-0.5em}
\end{figure}%

\begin{figure*}

  \centering
  \includegraphics[width=1.01\linewidth]{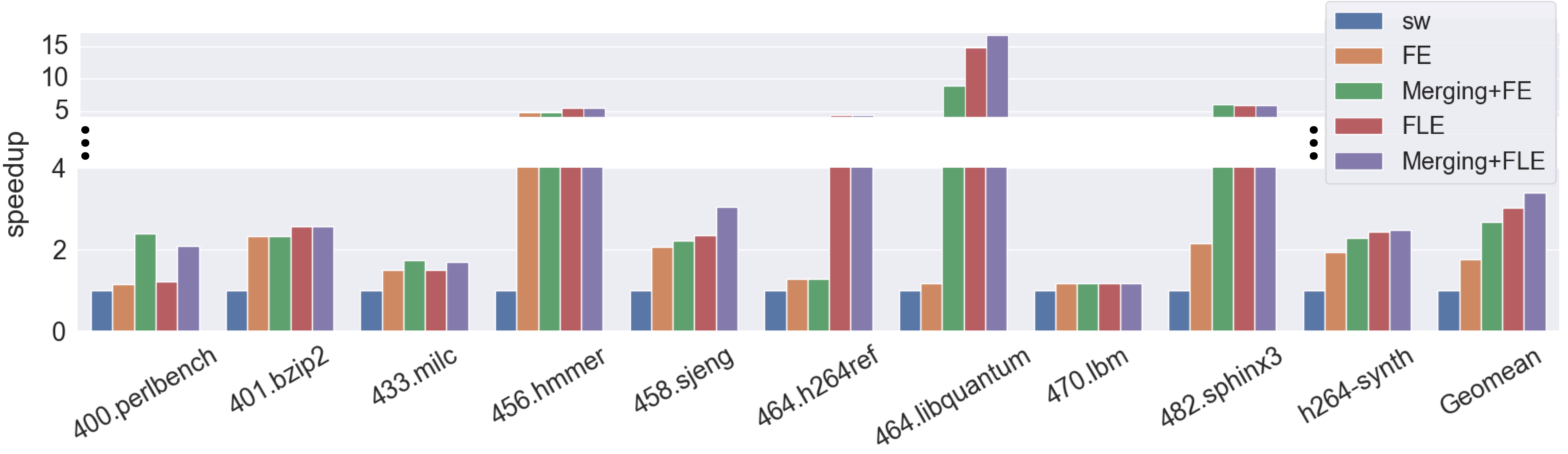}
  \caption{Overall speedup for SPEC applications \cite{henning2006spec} used in the FMSA~\cite{rocha2019function} and the Illinois \texttt{H.264}~\cite{h264uiuc}, for a 25 cycles interconnect latency and an area budget corresponding to the Artix Z-7007S used in DSE related work~\cite{zach2019compiler}. The "sw" bars represent the software baseline where no acceleration is being employed. The other configurations are described at the beginning of this section. Note the discontinuity in the y-axis used to represent the largest Speedups in the range $5 \times$ to $16.7 \times$. In this range the y-axes are more compressed than in the lower part of the figure.}
  \label{fig:overallSpeedup}
   \vspace{-0.1em}
\end{figure*}%

\begin{figure}
   \vspace{-0.5em}
  \centering
  \includegraphics[width=0.7\linewidth]{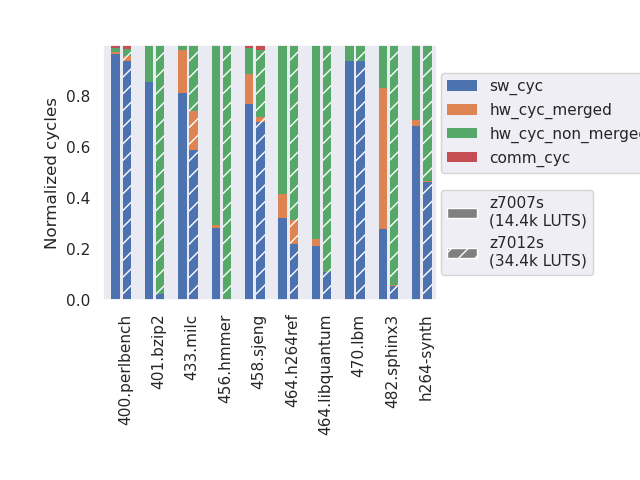}
    \vspace{-2.5em}
  \caption{Overall execution time cost breakdown for 25 cycles interconnect latency and an area budget corresponding to the Artix Z-7007S and Z-7012S used in AccelSeeker. All the applications in this plot are executed with the "FLE+Merging" configuration.}
  \label{fig:overallBreakdown}
  \vspace{-0.5em}
\end{figure}%

\textbf {Using MIP for HW/SW partitioning}: In this paper we use the Python-MIP~\cite{toffolo_santos} mixed integer programming package to find solutions to the HW/SW partitioning problem.Using MIP solvers for Early-DSE has the advantage of finding globally optimal solutions with the aid of high-performance libraries that exploit modern architectures. We use the Python-MIP package built on some of the fastest open source solvers, specifically the COIN-OR Branch-and-Cut solvers (CLP-CBC)~\cite{hans_mittelmann}. 


\textbf{Early DSE for accelerator merging faster than using Late-DSE exclusively}: Figure~\ref{fig:aladdinDSE} shows how the number of merging candidates to evaluate latency-wise would grow quadratically with the number of available input accelerators if we were to take a brute-force approach to coarse-grained accelerator merging and if every function were synthesizable. With as little as 40 input functions and loops, Aladdin takes over 15 hours of analysis and HLS takes about 5 hours to synthesize all the possible merged accelerators. Moreover, for non-synthesizable applications like the SPEC CPU2006 benchmarks we consider in this work, cycle accurate simulation and high level synthesis cannot even be applied, due to their expectation of a subset of the C language with explicit, unambiguous memory accesses and their lack of support for sophisticated computations with side-effects and recursion. AccelMerger is always able to produce useful SoC design insights without having to modify the application. The analysis time is short, as illustrated by Figure~\ref{fig:accelMergerDSE}. The increased variability for larger numbers of accelerable functions to merge is due to the fact that for a problem with more variables, it is not always the case that the problem takes longer to execute, since ILP optimization methods can immediately identify obvious candidates depending on the area budget and the software and hardware characteristics of the analyzed application.

Objective
\begin{multline}
  \label{eqn:obj}
  \quad
  \min_{\mathclap{\substack{hwv_{i},\\ swv_{i},\\ frontier_{ij} \in \{0, 1\} }}}
     \big(\sum_{1 \le i \le |P'|} (hwv_{i} \cdot hw_i +  swv_i \cdot sw_i) +
     \sum_{j \in C_i} calls_{ij} \cdot frontier_{ij} \cdot latency\big)
\end{multline}

Constraints
\begin{equation}
    \label{eqn:max_area}
    \sum_{1 \le i \le |P'|} hwv_{i} \cdot area_{i} \leq area\_budget  \\
\end{equation}
\begin{equation}
    \label{eqn:appRealization}
    swv_i + hwv_i + \sum_{j \in Descend_i} swv_j + hwv_j = 1 \text{, }  \forall i \in Roots \\
\end{equation}
\begin{multline}
    \label{eqn:callAndMerge}
    hwv_i + 1 - hwv_j - \sum_{k \in Descend_j} hwv_k \le 1  \forall i \in \{1..|P'|\} \text{, } j \in C_i \cap Roots
\end{multline}
\begin{equation}
    \label{eqn:comm}
    swv_i + hwv_i - frontier_{ij} \le 1 \: \text{, } \forall i \in \{1..|P'|\} \text{, } j \in C_i
\end{equation}

%% file: experiments.tex
\label{sec:setup}

\begin{figure*}
  \centering
  \includegraphics[width=1.01\linewidth]{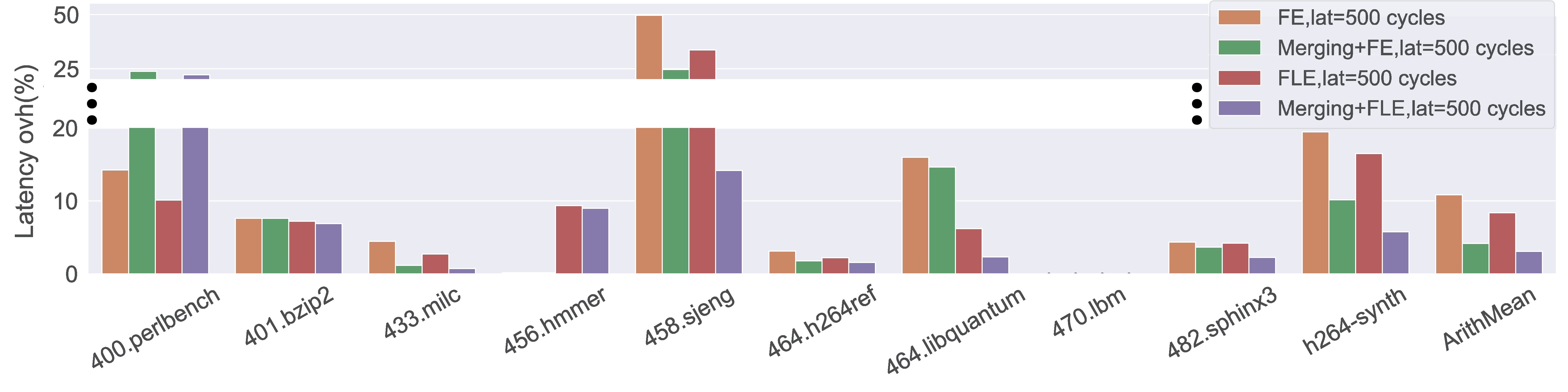}
  \caption{The percentage of application execution time spent in initiating and terminating accelerator computations in a large-latency situation and different merging and granularity scenarios. Note the discontinuity in the y-axis used to represent the largest Speedups in the range $25 \%$ to $50 \%$. In this range the y-axes are more compressed than in the lower part of the figure.}
  \label{fig:highLatImpact}
  \vspace{-0.5em}
\end{figure*}%

\subsection{Experimental Setup}
Within the scope of our experimental setup we use four configurations \textbf{FE}, \textbf{FLE}, \textbf{Merging+FE} and \textbf{Merging+FLE}. \textbf{Function Extraction (FE)} corresponds to using the Frontend and the HW/SW partitioning step in Figure~\ref{fig:workflow} to select functions for hardware realization. This configuration performs state-of-the-art early-stage accelerator selection~\cite{zach2019compiler}. The following three configurations exploit variable granularity and accelerator merging and represent AccelMerger's contribution. \textbf{Function and Loop Extraction (FLE)} is similar to \textbf{FE} but including the possibility of extracting loops as functions when deemed profitable. This configuration transforms the code to be more programmer independent in terms of how the code is split into functions and exploits the benefits of diversified granularity level (both fine and coarse-grained). \textbf{Merging + FE} and \textbf{Merging + FLE}  are similar to \textbf{FE} and \textbf{FLE} respectively, but they perform the intermediate step of accelerator merging from Figure~\ref{fig:workflow} as well as merging-aware HW/SW partitioning. 


\subsection{Overall performance improvement}

In Figure~\ref{fig:overallSpeedup} we showcase the benefits of accelerator merging when placing both the merged accelerators and the non-merged ones in the context of the whole application. 

A variety of applications yield different behavior as a result of varied granularities produced via merging and loop extraction. For example, performing loop extraction for the \texttt{milc} application shows almost no benefit (FLE versus FE). However the larger granularity of multiple merged functions is profitable since FE+Merging and FLE+Merging bring similar performance benefits, of $1.13\times$ with respect to FE and FLE and of $1.7\times$ with respect to the software version. A similar effect is occurring in the application, but with even stronger speedups for the configurations with merging. In \texttt{perlbench} we observe that FE+Merging is even better than FLE+Merging. Extracting loops into functions introduces too much overhead, a sign that functions are encapsulated well in \texttt{perlbench}.


\texttt{sjeng} in the Merging+FE configuration, brings smaller benefits than merging after loop extraction even though FLE does not bring significant benefits over FE, raising the speedup from $2.36\times$ for FLE to $3.04\times$ for the FLE+Merging configuration. This indicates that an intermediate granularity between function and loop level is the most suitable. On average we observe that the best configuration is FLE+Merging, reaching $3.39 \times$ with respect to software executions and $1.91 \times$ w.r.t. the FE configuration.

Figure~\ref{fig:overallBreakdown} shows that some applications (\texttt{perlbench}, \texttt{sjeng}, \texttt{milc} and \texttt{lbm}) spend most execution time in software. Other applications, like \texttt{libquantum}, \texttt{sphinx3}, or \texttt{hmmer} can mostly be realized in hardware at these large area budgets.  Since the hardware-software partitioning step is able to pick either the merged or the non-merged version of a function, it frequently picks the original functions at larger area budgets. Although they are more costly to realize in hardware, they are faster than merged counterparts burdened with multiplexing overhead. This effect is most dramatic in the case of \texttt{sphinx3}, followed by \texttt{libquantum} and \texttt{sjeng}.  However in \texttt{sjeng}, since there is much more left to realize in hardware than what is possible at these particular area budgets, we will see in Section~\ref{sec:otherAreaBud} that the benefit gained through merging fluctuates through the range of area budgets.

\begin{figure}
  \centering
  \vspace{-0.5em}
  \includegraphics[width=0.7\linewidth]{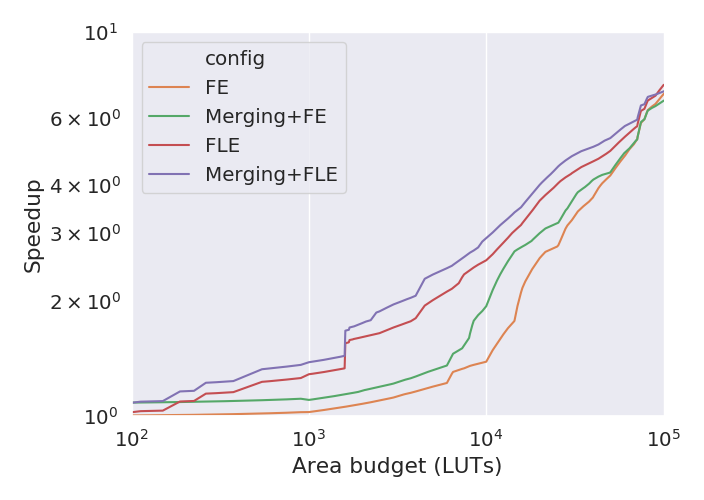}
  \caption{Speedup Geomean w.r.t. software execution across all applications for the FE, FE+Merging, FLE and FLE+Merging configuration under a wide range of area budgets. }
  \label{fig:fe_summary}
  \vspace{-0.5em}
\end{figure}

\subsection{Accelerator invocation latency impact}
\label{sec:comm}
A number of factors can hurt
the performance attainable via acceleration. Applications with high numbers of calls, for the most significant functions, tend to suffer from communication costs, since for each invocation of an accelerator, the host needs to set up memory-mapped model-specific registers (MSRs), and the accelerator notifies the host when the accelerated function and all its callees have finished. Beyond the large gap between memory access time and compute~\cite{memwall}, and  system configurations with off-chip FPGAs, long cache flushes are required before offloading computation onto accelerators~\cite{shao2016co}.
 
Any of these scenarios can lead to high latency in initiating communication with the accelerator. In Figure~\ref{fig:highLatImpact}, we show the maximum overhead represented by communication across all area, merging, and granularity configurations in such a high-latency scenario (500 cycles).

In general, we observe a trend for the FLE+Merging configuration to have slightly smaller overhead compared to the other configurations. 
For many applications the HW/SW partitioning algorithm is able to pick acceleration candidates with low overall latency, but for \texttt{sjeng}, \texttt{perlbench}, \texttt{H.264} and \texttt{libquantum}, we see that the worst communication latencies can be higher than 10\% and as high as 51.36\%. Conversely, applications with few calls to the most representative functions are impacted the least by latency costs. For example, \texttt{lbm} contains a single call to the most representative function \emph{LBM\_performStreamCollide}.

We see that for most area budgets and all latencies, merging has a beneficial impact in mitigating latency. The merged accelerators are able to cover more functions from the original application and therefore usually there is less switching between CPU computation and accelerator computation.





\subsection{Exploring other area budgets}
\label{sec:otherAreaBud}


In Figure~\ref{fig:fe_summary} we see what speedups can be accomplished with different area budgets using the state of the art in hardware-software partitioning as well as the three new techniques we introduce in this paper. This graph represents the average over all the applications considered in this work. For very small area budgets most of the applications run in software and thus the speedups are very close to 1. Also, most applications have a transition region where more and more functionalities can be realized in hardware, up to an area budget point when the whole application can be executed in hardware. Most applications start this transitioning region around 1000 LUTs, and all of them end it by 1M LUTs. When the area-latency curve has converged for a given application, that's equivalent to a monolithic accelerator scenario. \texttt{perlbench} is the most difficult application to fully realize in hardware. \texttt{hmmer} is the easiest and the least compelling for area-compressing techniques such as AccelMerger.

\begin{figure}
  \centering
  \includegraphics[width=0.7\linewidth]{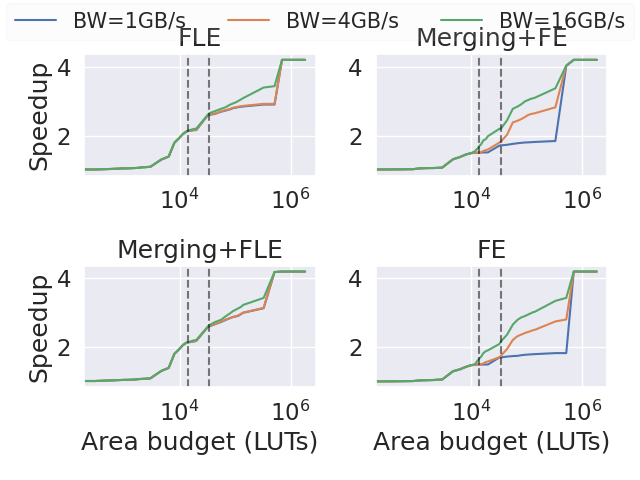}
  \caption{Bandwidth scenarios for varying granularities and merging configurations with the synthesizable version of \texttt{H.264}~\cite{h264uiuc}.}
  \label{fig:band_h264}
   \vspace{-0.5em}
\end{figure}

Figure~\ref{fig:fe_summary} contains a superset of the results shown in Figure~\ref{fig:overallSpeedup} for the Geomean bars. The average speedups for the Merging+FE and Merging+FLE is driven by the applications \texttt{milc}, \texttt{sphinx}, \texttt{H.264}, \texttt{H.264ref}, \texttt{sjeng} and \texttt{perlbench} which have significant performance improvements over the FE configuration for some considerable area budget ranges. Another observation is that in order to extract the maximum speedup from the applications, working with the original application functions limits the maximum achievable speedups, thus encouraging the FLE and Merging+FLE configurations. 




\subsection{Bandwidth use-case for \texttt{H.264}}


For the synthesizable implementation of the \texttt{H.264} decoder~\cite{h264uiuc}, Figure~\ref{fig:band_h264} shows the effects on performance of varying bandwidth over a wide range of area budgets. For the FE configuration, as expected, more bandwidth yields more speedup. For FLE, loop extraction shows greater low bandwidth tolerance for bandwidths of $1GB/s$ and $4GB/s$. Increasing the pool of available functions, with varying communication requirements, opens fresh hardware acceleration opportunities. The addition of merging extends tolerance to lower bandwidths by making more functions available. Moreover, at high area budgets, merging involves less switching between accelerators and the CPU, so less data movement is required.

%% file: conclusion.tex
Early stage accelerator design through function merging, based on optimized selection of merged and non-merged, hardware-realized functions and loops, opens an exciting research area that promises to benefit performance and area/latency trade-offs. AccelMerger can enable lower, system-level data communication, with a focus on interconnect latency. It allows designers to explore accelerators of varying granularities without deep application knowledge, as the choice of accelerators is not fixed by the structure of an application's source code. Experimental results show up to $16.7\times$ performance improvement over software-only implementations, and $1.91\times$ on average over state-of-the-art HW/SW  partitioning tools. In future work, we plan to extend AccelMerger's SoC-level communication analysis, apply it for cross-application merges and add power models to its repertoire. 